\documentclass{article}
\usepackage{url}
\usepackage[pdftex,plainpages=false,colorlinks,citecolor=black,linkcolor=black,urlcolor=black,bookmarksopen]{hyperref}
\usepackage{csquotes}
\usepackage{url}

\usepackage[backend=biber,style=numeric,sorting=none,isbn=false,url=false,date=year,clearlang=true]{biblatex}
\AtEveryBibitem{\clearlist{language}}
\DeclareDelimFormat{nameyeardelim}{\addcomma\space}

\DefineBibliographyStrings{english}{
  andothers = {\mkbibemph{et\addabbrvspace al\adddot}}
}
\usepackage{longtable}
\usepackage[pdftex]{graphicx}
\usepackage{subfigure}
\usepackage{pdfpages}
\usepackage{titlesec}
\usepackage{booktabs}
\usepackage{fontawesome}
\usepackage{setspace}
\usepackage{parskip} 
\usepackage[a4paper,layout=a4paper,
        left=2.5cm,right=2.5cm,
        head=2.5cm, bmargin=2.5cm]{geometry}
\usepackage{cleveref}
\usepackage{multirow}
\usepackage{makecell}
\usepackage{colortbl}
\usepackage{enumitem}
\usepackage{placeins}
\usepackage{tabularx}
\usepackage{siunitx}
\usepackage{subcaption}
\usepackage{lscape}
\usepackage{array}
\usepackage{multicol}
\newcolumntype{o}{X}
\newcolumntype{s}{>{\hsize=.6\hsize}X}
\newcolumntype{Y}{>{\centering\arraybackslash}p{0.025\textwidth}}
\usepackage{tikz}
\usetikzlibrary{matrix, positioning}
\providecommand{\keywords}[1]
{
  \small	
  \textbf{\hspace{2em}\textit{Keywords---}} #1
}

\newcolumntype{C}[1]{>{\centering\arraybackslash\hspace{0pt}}p{#1}}

\newcommand{\sph}[2][13em]{
	\rotatebox{90}{\parbox{#1}{\raggedright #2}}}

\newcommand*\hc{\tikz\draw[fill] (0,0)-- (90:1ex) arc (90:270:1ex) -- cycle;} 
\newcommand*\fc{\tikz\fill (0,0) circle (1.0ex);} 

\newcolumntype{u}{p{0.0048\textwidth}}

\usepackage{fancyhdr}
\title{Designing an Interdisciplinary Artificial Intelligence Curriculum for Engineering: Evaluation and Insights from Experts}
\author{Johannes Schleiss \and Anke Manukjan \and Michelle Ines Bieber \and Sebastian Lang \and Sebastian Stober}
\date{\small{Otto von Guericke University Magdeburg, Germany}}

\begin{document}

\maketitle
\fancyhead[L]{\textit{Designing an AI Curriculum for Engineering}} 
\fancyhead[C]{}                            
\fancyhead[R]{\thepage} 

\fancyfoot[L]{}                    
\fancyfoot[C]{}                            
\fancyfoot[R]{}                            


\begin{abstract}
As Artificial Intelligence (AI) increasingly impacts professional practice, there is a growing need to AI-related competencies into higher education curricula. 
However, research on the implementation of AI education within study programs remains limited and requires new forms of collaboration across disciplines. 
This study addresses this gap and explores perspectives on interdisciplinary curriculum development through the lens of different stakeholders. 
In particular, we examine the case of curriculum development for a novel undergraduate program in AI in engineering. 
The research uses a mixed methods approach, combining quantitative curriculum mapping with qualitative focus group interviews. 
In addition to assessing the alignment of the curriculum with the targeted competencies, the study also examines the perceived quality, consistency, practicality and effectiveness from both academic and industry perspectives, as well as differences in perceptions between educators who were involved in the development and those who were not.
The findings provide a practical understanding of the outcomes of interdisciplinary AI curriculum development and contribute to a broader understanding of how educator participation in curriculum development influences perceptions of quality aspects. 
It also advances the field of AI education by providing a reference point and insights for further interdisciplinary curriculum developments in response to evolving industry needs.
\end{abstract}
\keywords{Curriculum Design, Engineering Education, Program evaluation, Interdisciplinary groups, Artificial Intelligence in Engineering}

\section{Introduction}
\label{sec:introduction}
The integration of data and technology in education increases the demand for AI education and respective curriculum changes~\cite{gurdur_broo_rethinking_2022,patel_artificial_2021}. 
AI impacts education from two perspectives.
On the one hand, using AI tools in education can open up new teaching and learning methodologies~\cite{bond2024meta,johri2023generative}. 
On the other hand, the use of AI tools in professional context forms new roles and required competencies at the intersection of AI and the domain, making it necessary to integrate these topics into the curriculum~\cite{chiu2024future}.

Various teaching approaches for AI exist, targeting a wide range of audiences~\cite{almatrafi_systematic_2024,druga2022landscape,laupichler2022artificial}.
While K-12 curricula have received considerable attention~\cite{chiu2021creation,kim2021and,casal2023ai,yang2022artificial,su2022meta}, interdisciplinary and domain-specific AI education remains underexplored in terms of competencies, courses and curricula~\cite{knoth_developing_2024,chiu2024future}.

This paper contributes to enhancing this understanding of developing interdisciplinary AI curriculum by presenting and analyzing a case study of the development of an interdisciplinary curriculum for a Bachelor's program at the intersection of AI and engineering.
In particular, we focus on translating an interdisciplinary competence profile for the application of AI in engineering~\cite{schleiss2022interdisciplinary} into an interdisciplinary curriculum program.
Although computing and computing education have already become an integral component of the engineering profession and engineering education and topics such as programming, data analysis, and computational sciences are now commonly taught in engineering curricula \cite{malmi_selective_2023,raj_empirical_2019}, there exist limited experiences and systematic insights into teaching AI in an engineering context.
During the last few years, engineering educators have integrated AI education in single lectures, courses, or projects~\cite{isaac_flores-alonso_introduction_2023,ng_review_2023,singelmann_framework_2023}.
Moreover, the necessity to develop AI competencies is recognized in across disciplines~\cite{pinski2024ai}.
At the same time, the integration of education about AI on a curricular level or developing a full interdisciplinary study program in higher education from has not received much focus.
This connects to the overall limited focus on curriculum scholarship in higher education~\cite{krause2022vectors}.
Consequently, the understanding of developing and implementing such novel and interdisciplinary curricula is still limited.

The study employs a mixed-method approach based on a case study with quantitively validating the expected effectiveness through a curriculum mapping from the intended competence profile to the courses and qualitative focus group interviews with lecturers and industry experts.
Overall, the study builds a reference for an interdisciplinary AI curriculum for engineering, highlighting the development and evaluation of it and improving the understanding of how the participation of educators in curriculum development changes their perception of quality aspects. 
It study contributes to understanding aspects of interdisciplinary AI curriculum development with a focus on AI and engineering.

\subsection{Curriculum Development and Evaluation}
The concept of the curriculum has different definitions and conceptions \cite{lunenburg2011theorizing}.
In the context of this study, we refer to a curriculum as the overall plan of modules and student experiences in an educational program, following the short definition of Taba \textcite{taba1962curriculum} as a \enquote{plan for learning}. 

\paragraph{Curriculum Development}
Following Ornstein and Hunkins the idea of curriculum development is \enquote{to show how [a] curriculum is planned, implemented, and evaluated as well as what people, processes, and procedures are involved in constructing the curriculum}~\cite[p. 30]{ornstein2017curriculum}.
Scholars have proposed various curriculum development frameworks, each with its emphasis and underlying assumptions.
Following \textcite{visscher2004paradigms}, four main approaches exist:
The \emph{instrumental approach} follows systematic, linear planning with defined goals.
The \emph{communicative approach} emphasizes stakeholder participation and consensus-building whereas the \emph{artistic approach} treats curriculum development as a creative process.
Last, the \emph{pragmatic approach} prioritizes iterative development and real-world feedback.
In practice, the approaches are often combined as illustrated by \textcite{schaper2012fachgutachten} three development paths for higher education.
The authors distinguish between (1) the use of existing guiding principles and educational standards or competency profiles, (2) surveying graduates of comparable study programs and subject-specific employers, and (3) participatory methods for the development of novel and non-comparable degree programs.
This development approach combines participatory methods with structured frameworks like Backward Design~\cite{wiggins2005understanding} to achieve a balanced solution that maintains rigor while accommodating stakeholder input and needs.

\paragraph{Curriculum Evaluation}
Assessing curriculum quality is complex and involves two types: formative and summative evaluation~\cite{nieveen2013formative} and can focus on assessing different forms of a curriculum.
\textcite{thijs2009curriculum} highlighted three main forms of a curriculum:
The \emph{intended curriculum} refers to the curriculum designed by the curriculum developers.
Second, the \emph{implemented curriculum} is the curriculum that the educators implement in practice.
Third, the \emph{attained curriculum} refers to perceived learning outcomes of learners.
This study focuses on the intended curriculum, analyzing the developed curriculum of the case study using a formative evaluation.
Following ~\textcite{nieveen2013formative}'s quality criteria, we evaluate
(1) \emph{relevance} (intervention is needed and based on latest research), (2) \emph{consistency} (curriculum structure is logical and cohesive), (3) \emph{expected practicality} (anticipated ease and readiness of use), and (4) \emph{expected effectiveness} (anticipated achievement of desired outcomes).
The actual practicality and effectiveness can only be assessed for the attained curriculum after its implementation and is not emphasized here.
 
\paragraph{Methods of Formative Evaluation}
Various formative evaluation methods have been proposed and the choice of method varies by design phase~\cite{thijs2009curriculum,nieveen2013formative}. 
This study uses the method of screening through curriculum mapping and expert reviews through focus group interviews with educators and industry professionals.

In the context of this study, we focus on the following research questions: 

\textit{RQ1: To what extent does the curriculum align with the targeted competence profile for AI in engineering? 
}

\textit{RQ2: How do educators and industry representatives perceive the quality, consistency, practicality, effectiveness, and interdisciplinarity of the curriculum?
}  

\subsection{Interdisciplinary Curriculum Development and Educators' Perspectives}
With more curriculum change efforts needed in the light of AI education, it is necessary to understand the practical aspect of interdisciplinary curriculum development to include a novel set of competencies of AI or create novel curriculum approaches.
Interdisciplinary education aims to bridge different epistemologies and integrate content and concepts from different disciplines into one teaching approach \cite{van2020interdisciplinary,spelt_teaching_2009,lindvig_different_2019}.
Effective interdisciplinary curriculum design requires a cohesive approach, yet the degree of integration can vary significantly, leading to strong or weak interdisciplinarity \cite{augsburg2009politics,klein2010creating,knight2013understanding}. 
Strong programs often feature core interdisciplinary courses and dedicated faculty, whereas weak programs allow students to choose courses, which are not intentionally integrated, leaving the integration across subjects to the students themselves \cite{knight2013understanding}.
However, the impact on student learning of these features remains still empirically not clear~\cite{lattuca2017examining}.

Curriculum development is influenced by socio-cultural factors, including educators' beliefs and their perspectives on their disciplines~\cite{lattuca2009shaping,stark_faculty_1988,lattuca2016towards}.
If educators are part of the curriculum development of programs, their strong beliefs about their own discipline make interdisciplinary approaches more challenging highlighting the need for communicative approaches that allow for deliberations.
At the same time, participating in the development process educators might advance a more holistic picture with less strong views on the discipline.

To move towards some qualitative evidence, we explore the perceived quality of educators involved in development and those who were not, hypothesizing that participation fosters broader perspectives. 

\textit{RQ3: What are the differences in perceptions and expectations among industry representatives, participating educators, and non-participating educators regarding the curriculum?}

The remainder of the paper is structured as follows:
\Cref{sec:methods} focuses on the underlying methodology including the research approach, providing context for the case study and the evaluation instruments of curriculum mapping and focus groups.
\Cref{sec:evaluation} presents the results from the curriculum mapping and the qualitative insights from the focus group interviews.
The results are discussed and contextualized in \Cref{sec:discussion}. 
\Cref{sec:conclusion} summarizes the main findings and provides an outlook on future work.

\section{Methods and Materials}
\label{sec:methods}
\subsection{Methods}
The study follows a case study methodology~\cite{yin2018case} to evaluate and enhance an interdisciplinary curriculum, specifically focusing on a novel bachelor program at the intersection of AI and engineering. 
It focuses on the intended curriculum, emphasizing quality assurance in the program development and gaining understanding of educator's perspectives.
To analyze the curriculum, we used a mixed-method approach with a quantitative aspect of curriculum mapping and a qualitative aspect of focus group interviews with lecturers and industry experts. 

\subsection{Case-Study Bachelor Program AI Engineering}
The program operates cooperatively across five universities in a regional state in [Blinded]. 
Its interdisciplinary nature was fostered through participatory curriculum workshops, which engaged various stakeholders in the development process~\cite{schleiss2023curriculum,bieber2023curriculumwerkstatt,manukjan2023curriculumwerkstatt}.
To contextualize the case study, we briefly describe the underlying developed competence profile and curriculum.

\paragraph{Competence Profile}
The competence profile describes the competencies students should possess after completing the program. 
Developed collaboratively among stakeholders, it reflects the latest research results, educators and industry perspectives~\cite{schleiss2022interdisciplinary,schleiss2023curriculum}.
The targeted competence profile of the program can be distinguished into categories of math, engineering, computer science, AI, engineering specialization, process \& system-oriented working, and interdisciplinary skills (\Cref{tab:competence_profile_ai_eng}).

\begin{figure}
\centering
	\includegraphics[width=\textwidth]{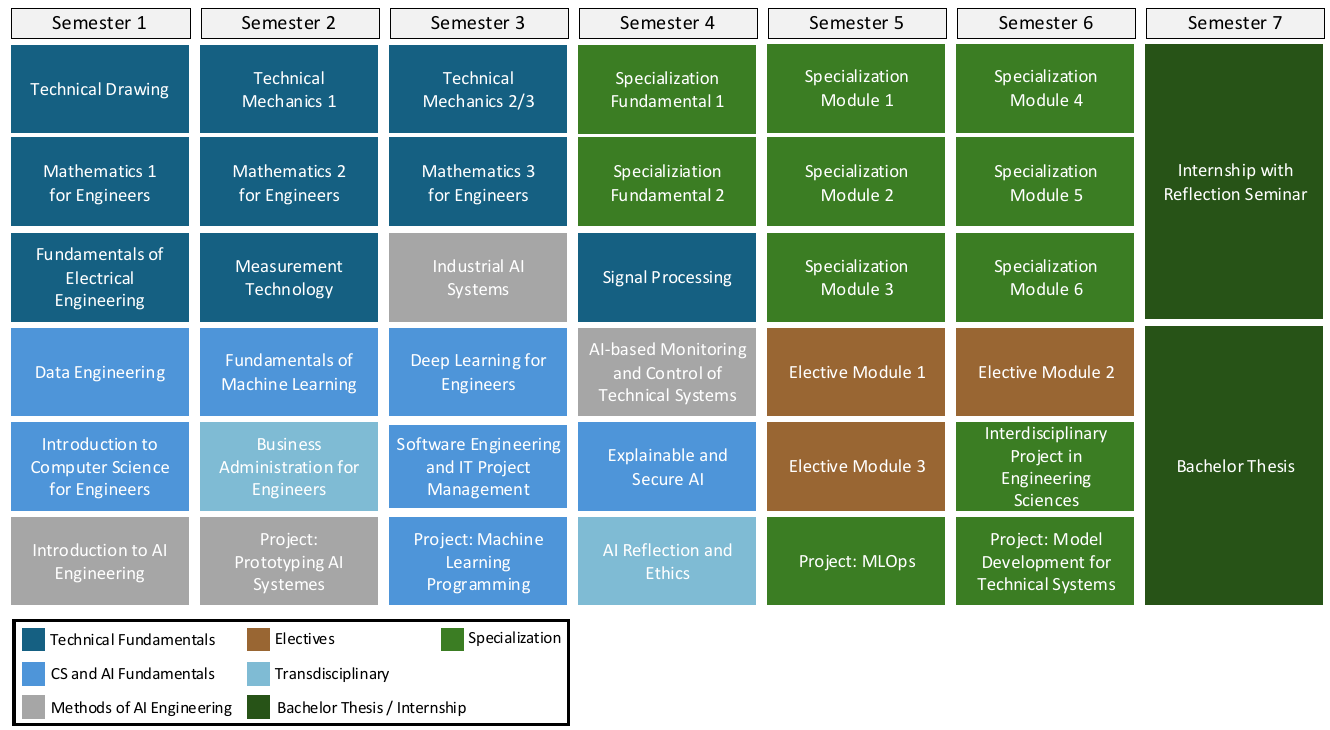}
	\caption[Curriculum of the AI Engineering study program over seven semesters.]{Curriculum of the study program AI Engineering over seven semesters. Each module has a load of 5 credit points, corresponding to 150 hours.}
	\label{fig:curriculum_core}
\end{figure}

\paragraph{Curriculum}
According to the backward design, the program-level competency profile was used to develop a curriculum to achieve the targeted competencies. 
The seven-semester curriculum (210 credits points (CP)) is divided into two parts (visualized in \Cref{fig:curriculum_core}). 
First, students take a core curriculum in semesters 1-4 that provides a foundation in engineering, math, computer science, and AI, as well as focus on integrative courses and projects that bring engineering and AI together.


After completing the core curriculum, students choose one of five specializations that relate to engineering application areas.
In the specializations, students focus on using AI technology in the context of a particular application area, emphasizing working with data and engineering processes from the domain and understanding the underlying foundations of the application area.
Five different specializations currently offered in this curriculum are: (1) agricultural economy and technology, (2) biomechanics and smart health technologies, (3) green engineering, (4) manufacturing, production and logistics, and (5) mobile systems and telematics.
The curricula of all specializations are shown in \Cref{appendix:curricular_details} (\Cref{fig:appendix_curriculum_specialization_1}, \Cref{fig:appendix_curriculum_specialization_2}, \Cref{fig:appendix_curriculum_specialization_3}, \Cref{fig:appendix_curriculum_specialization_4}, \Cref{fig:appendix_curriculum_specialization_5}). 

\subsection{Evaluation Instruments}
\label{subsec:eval_instruments}
To evaluate the developed curriculum, two formative evaluation instruments were used: curriculum outcome mapping and focus group interviews with experts, following~\textcite{thijs2009curriculum}.

\paragraph{Curriculum Outcome Mapping}
This approach assesses the alignment between the intended competencies in the competency profile and the learning outcomes of the curriculum modules~\cite{allen2003assessing}.
One researcher mapped the competencies to the curriculum modules based on descriptions from the module catalog (categories: partly, fully or not addressed).
To ensure reliability, a second person with a different background coded parts of the mapping and divergent codes were discussed.
The mapping focused only on learning objectives and content descriptions, minimizing biases from raters.
After coding, scores for each competency sub-area were summed to assess overall coverage, and averages for each competency group (e.g., math, engineering fundamentals) were calculated to highlight areas of emphasis.

\paragraph{Focus Group Interviews}
To deepen the understanding of curriculum mapping, focus group interviews were conducted with educators who teach in the curriculum, and industry experts.
Focus group interviews gather a small, diverse group of people, usually experts in a field, to discuss a specific topic or set of questions, allowing researchers to observe group interactions and gather a range of perspectives~\cite{kitzinger1995qualitative,morgan1996focus}. 
The main reason for using focus group discussions is to generate debate and collective insights on a research topic, revealing the underlying meanings, experiences, and beliefs~\cite[p.~28]{o2018use}. 
Participants were selected for their relevance to the topic, age, social characteristics, and comfort with discussion, adding depth and richness to the data through group dynamics~\cite[p.~655]{rabiee2004focus}.

A total of six interviews were conducted with 19 people across three groups (two interviews per group). 
The groups differed as (1) educators who were part of the curriculum development, (2) educators who were not part of the development, and (3) industry experts.
The distinction between the groups was intended to avoid positive and negative bias and to assess how the perceptions of the groups differed. 

The interviews followed a clear structured guide to allow for comparisons between interviews. 
The interview guide can be found in \Cref{appendix:interview_guide}.
The interviews lasted approximately one hour each and were conducted by two researchers, one as moderator and the other as observer and note taker.
In addition, they were audio-recorded and transcribed. 

The coding scheme was developed inductively and iteratively per question using qualitative content analysis~\cite{mayring2004qualitative} and stepwise thematic analysis~\cite{braun2006using}.
To ensure a reliable coding process, the interviews were then paraphrased.
In the coding phase, each of the three participating researchers coded a subsets of the semi-structured interviews to ensure consistency across interviews.
To ensure reliability of coding, a second coder coded one-third of the interviews, and disagreements were discussed and resolved.
The resulting coding scheme is presented in \Cref{tab:coding_scheme}.
The data was summarized based on the different interview sections and codes, and compared between the different focus groups of industry, participating, and non-participating educators.
In addition, key quotations were marked and included in the results.

\section{Results}
\label{sec:evaluation}
In the following, the results of the formative evaluation are reported, including the curriculum mapping and the qualitative analysis through a focus group interview with experts. 

\subsection{Curriculum Outcome Mapping}
Curriculum outcome mapping was used as a checkpoint to ensure a consistent and practical design and to assess whether the modules could achieve the desired competency profile.

\paragraph{Descriptive Elements}
The total credit load for the program is 210 credit points (CP).
In addition to the internship and the Bachelor thesis, a total of five explicit project modules (25 CP) have been identified.
Three modules (15 CP) are electives and were not included in the curriculum mapping.
The curriculum consists of a core curriculum (110 CP) and a chosen specialization (100 CP). 

\paragraph{Program Level Coverage of Competencies}
In the mapping, the subdimensions of all competency areas were coded on three levels (\textit{0 - not covered}, \textit{0.5 - partially covered}, and \textit{1 - covered}).
To understand the coverage and focus, the summed scores per competency area and the average of the summed subdimensions for the core curriculum and specializations are reported in \Cref{tab:curriculum_mapping_coverage}.
In addition, the detailed mapping, including all courses and competency dimensions, can be found in \Cref{appendix_modulmapping} \Cref{tab:curriculum_mapping}.


\FloatBarrier
\begin{table}[!htpb]
	\centering
	\caption[Coverage of competency areas after core curriculum and after each specialization.]{Coverage of competency areas after core curriculum and after each specialization. Per area (Math (M), Engineering (Eng), Computer Science (CS), Artificial Intelligence (AI), Engineering Specialization (ES), Process- and Systems Thinking and Working with AI (PS), Transversal (T)),  the sum $\sum$ and the average of the mapped sub-competencies $\overline{c}$ are reported to understand the emphasis and coverage of the competencies. The complete mapping can be found in the appendix~\Cref{appendix_modulmapping}.}
	\label{tab:curriculum_mapping_coverage}
	\footnotesize
	\begin{tabular}{p{0.25\linewidth} YYYYYYYYYYYYYY}
		\toprule 
	
	&
		\multicolumn{2}{Y}{\textbf{M}} &
\multicolumn{2}{Y}{\textbf{Eng}} &
		\multicolumn{2}{Y}{\textbf{CS}} &
		\multicolumn{2}{Y}{\textbf{AI}} &
		\multicolumn{2}{Y}{\textbf{ES}} &
		\multicolumn{2}{Y}{\textbf{PS}} &
		\multicolumn{2}{Y}{\textbf{T}} \\
		\cline{2-3} \cline{4-5} \cline{6-7} \cline{8-9} \cline{10-11}  \cline{12-13}  \cline{14-15}
		& $\sum$ & $\overline{c}$ & $\sum$  & $\overline{c}$ & $\sum$ & $\overline{c}$ & $\sum$  & $\overline{c}$ & $\sum$ & $\overline{c}$ & $\sum$  & $\overline{c}$ & $\sum$  & $\overline{c}$ \\
		\midrule
		Core Curriculum                                  						  & 5.5 & 0.9  & \phantom{0}9.5  & 1.4  & 3.0 & 0.8  & 18.5 & 2.6  & 1.0 & 0.5  & \phantom{0}6.5  & 2.2  & \phantom{0}5.0  & 1.3  \\
		S1:   Agricultural economy and technology         		& 5.5 & 0.9  & 10.5 & 1.5  & 4.0 & 1.0  & 27.5 & 3.9  & 8.0 & 4.0  & 11.0 & 3.7  & 13.0 & 3.3  \\
		S2: Biomechanics and smart health technologies 	  & 5.5 & 0.9  & 10.5 & 1.5  & 4.0 & 1.0  & 29.5 & 4.2  & 6.5 & 3.3  & 12.0 & 4.0  & 13.5 & 3.4  \\
		S3: Production, manufacturing and logistics    	   & 7.0 & 1.2  & 11.5 & 1.6  & 4.0 & 1.0  & 29.5 & 4.2  & 7.0 & 3.5  & 11.5 & 3.8  & 12.0 & 3.0  \\
		S4: Green engineering                          						  & 6.0 & 1.0  & 10.5 & 1.5  & 4.0 & 1.0  & 28.0 & 4.0  & 7.0 & 3.5  & 10.5 & 3.5  & 13.5 & 3.4  \\
		S5: Mobile systems and telematics                   		 & 5.5 & 0.9  & 11.0 & 1.6  & 4.0 & 1.0  & 28.5 & 4.1  & 8.5 & 4.3  & 11.5 & 3.8  & 12.0 & 3.0 \\
	\end{tabular}%
\end{table}

\paragraph{Coverage of Competencies}
The results indicate that the curriculum is aligned with the targeted competency profile, but the coverage varies across the different areas. 
As indicated by the average values $\overline{c}$ in \Cref{tab:curriculum_mapping_coverage}, there is a strong emphasis on AI-related competencies, engineering specializations, process/systems-oriented work with AI, and transversal skills.
However, there are also sub-competencies, particularly in math and fundamental machine learning topics, that are only partially covered, indicating that these topics are present but may not be thoroughly addressed in the curriculum learning outcomes or module descriptions.
In addition, there is limited explicit emphasis on the fundamentals of computer science and data science, and deeper reinforcement of these skills can only be explicitly identified in the project modules. 

The curriculum focuses more on AI system development and prototyping, with less coverage of the skills needed to deploy, monitor, and maintain AI systems in production.
For example, server and cloud application skills are only covered in the MLOps project module, and ethical considerations are only explicitly covered in the core curriculum.

Overall, the mapping shows that the curriculum follows a logical progression with no clearly identified gaps, with earlier modules focusing on basic skills and their integration, and later modules building on these to develop more advanced, specialized skills.
Integration and transfer occur mainly in the project modules and in specialized interdisciplinary modules.
\subsection{Qualitative Analysis Through Focus Groups}
\paragraph{Participants}
A total of 19 experts participated in the six focus group interviews. 
Of these, five were industry experts and 14 were educators. 
The educators were divided into two groups: eight participating educators who were involved in curriculum development and six non-participating educators.
All educators had at least four years of teaching experience, with the participating educators having slightly more experience on average.
Almost three-quarters of the educators had previous experience in curriculum development, while half of the non-participating educators had no such experience (n=3).
The industry experts had several years of professional experience, with most having at least seven years, and almost all working with AI on a daily basis.
Overall, almost all respondents had a very positive or positive attitude towards interdisciplinary programs in general, with only two educators expressing a neutral attitude.

\paragraph{Effectiveness and Consistency: Curriculum and Competency Profile Fit}
When asked about the effectiveness and consistency of the curriculum's fit with the competency profile, participants rated the fit overall as good, with detailed ratings provided in \Cref{tab:ratings_fit}.

\begin{table}[ht]
	\centering
	\caption[Reported fit of curriculum and competencies.]{Reported fit of curriculum and competencies per status group. Participants had to rate their initial response before discussing the question ``How well is the competency profile achieved with this curricular composition of the modules?". Ratings on a five-point Likert scale ranging from \textit{1-not good at all} to \textit{5-very good}. Number of participants per group N, mean of rating, and standard deviation (SD).}
	\begin{tabular}{l c c c}
		\toprule
		\textbf{Group} & \textbf{N} & \textbf{Mean} & \textbf{SD} \\ \midrule
		Participating educators  & 8 & 4.50 & 0.53\\
		Non-participating educators  & 6& 3.17 & 0.75\\
		Industry & 5 & 3.80 & 0.45\\
		\midrule
		Total & 19 & 3.89 & 0.89 \\
		\bottomrule
	\end{tabular}
	\label{tab:ratings_fit}
\end{table}

In terms of the reasons for their perceptions, participants criticized missing content (n=12) such as deployment and operations, preprocessing, and generative AI, as well as perceived gaps in the foundations of computer science.
There were also concerns about the true depth and application of the competencies, for example in the context of programming (n=9), and the challenge of integrating and linking domains (n=8).
Some felt that the implementation of certain competencies could be difficult given the curriculum load (n=6). 
Representing and teaching transferable skills in the curriculum was also seen as a challenge (n=6).
Participants also identified the implementation of the program in general (n=4) and providing students orientation (n=4) as challenging. 
For example, participants stated:

\textit{``[...] how well the bridge between engineering and AI can be built afterwards really depends on how this can really be implemented. [...] We have a bit of a challenge to keep the thread going from the fundamentals back here into the specialization.'' [PE5]}

\textit{``We will still see a lot during the implementation where there may be gaps and where the theoretical plans may not work. But from what we have been able to prepare now, it fits very well.'' [PE7]}

Throughout the interviews, overburdening of students was identified as a risk factor (n=8), mostly related to overburdening by interdisciplinarity and diversity of competencies for a Bachelor's program. 
On a positive note, participants highlighted the practical and project-oriented nature of the curriculum (n=4).

\paragraph{Practicality: Strength and Weaknesses}
To further understand the practicality of the program, participants were also asked what they perceived to be the strengths and weaknesses of the program.
Strengths mentioned were the practical and project orientation (n=5), the interdisciplinary orientation of the program (n=3), the curriculum in general (n=2), its diversity (n=2), and the potential employability (n=2).
Weaknesses mentioned were that the development could have been more innovative (n=3), that graduates might not be industry-ready after a Bachelor's degree (n=2), and single mentions of the high diversity as a risk of losing orientation, lack of foundation, and lack of time and opportunity to deepen skills.

\paragraph{Fit to Disciplinary Expectations}
To understand the diversity of disciplinary expectations, participants were asked to give a subjective rating of the curriculum's fit with their disciplinary expectations.
Because this subjective rating is not comparable due to different interpretations of expectations and the question, overall observations are reported.
Participants generally felt that certain parts of the curriculum lacked the depth expected in more traditional programs such as computer science or engineering.
A common concern was that there was a lack of foundational coverage in some areas, with too much breadth and not enough depth across the curriculum.
Specifically from a computer science perspective, participants felt that the modules may not provide enough programming experience.
While the interdisciplinary nature of the program was seen as a strength, it was also seen as a potential challenge in meeting disciplinary expectations and providing sufficient depth in the various areas covered.

\paragraph{Opportunities and Risks of Interdisciplinarity in the Program}
Participants also discussed the opportunities and risks of the interdisciplinary nature of the program. 
Opportunities mentioned included integration and linking of content (n=6), practice and project orientation (n=5), diversity of content (n=4), profiling (n=4), development of a holistic view (n=4), communication across disciplines (n=3), change of perspective (n=3), and improved employability (n=2).
As one industry participant noted, \textit{``the greatest opportunity of interdisciplinarity is that [...] graduates actually also have opportunities in their professional lives outside of these five specializations [...] because they have a background that is relatively universal.'' [I3]}

Risks included challenges in integrating content (n=11), potential overload because of interdisciplinarity (n=5), effort and cost of implementation (n=3), lack of depth (n=2), and overall difficulty of implementation (n=3).
One participant explained: \textit{``I see the risk that it really is an overloading demand. [...] Everyone has to think for themselves, how can I somehow link what I learn from measurement technology with the basics of machine learning, and I think that's actually a difficult transfer.'' [I2]}

Challenges included cooperation across domains in implementing the program (n=4), communication with the students in the program (n=2), marketing to the target group, and setting expectations for the program (n=2).

Participants viewed the integration of courses and domains as an opportunity (n=6), a risk (n=11), and a challenge (n=8). 
While integration was seen as beneficial, concerns remained about effective implementation and potential fragmentation of knowledge.

As one participant noted, \textit{``If it's all integrated really well and works, then I think it's a pretty cool program [...] The risk is that if there's an issue at any point, it can quickly turn out really bad for the students.'' [NPE3]}

\paragraph{Potential Challenges and Improvements}
Other challenges mentioned by the participants were the implementation of the program (n=6), especially regarding the integration and organization of interdisciplinary content and the question if educators can adapt to students. 
In addition, small student cohorts were seen as a challenge, especially in the context of implementing and filling the five specializations (n=4).
Giving orientation for students (n=4) and the practical orientation (n=3) were also identified as challenges.

In terms of program improvements, participants had several suggestions. Two participants recommended identifying core practices and competencies and emphasizing them in the curriculum. One participant each suggested increasing communication about the interplay of modules and courses, reviewing the sequencing of modules, changing the orientation of technical topics, moving ethical discussions to a later semester, and preparing students' communication skills.

As potential solutions, participants suggested supporting interlinking and integration (n=3) and collaborating on implementation (n=3).
This could be achieved through a meeting of educators who teach in the same semester or through improved documentation of the content accessible to all educators, i.e., providing information about the modules or the technologies and tools used (n=3) and providing information to students through additional events (n=1). 

\paragraph{Differences from Participation in the Curriculum Development}
Overall, the interviews showed that people who participated in the development of the program were able to contextualize the outcome of the development to a greater extent, and also to justify and defend the design choices made in the development, as observed in this interview excerpt:

\textit{``This is an interdisciplinary degree course. This means that there is always the necessity or the fact that we can't train a jack-of-all-trades in a degree program like this and that we have to cut back in certain areas. [...] We have also discussed what can be achieved in many workshops and it now fits in with what we wanted on paper.'' [PE7]} 

Positive highlights such as the practical orientation (participating educators (pe)=5), the interdisciplinary orientation (pe=3) and an optimized curriculum (pe=2) were only mentioned by participating educators.
However, participating educators also acknowledged the challenge of providing orientation for students (pe=4), marketing (pe=2), and the reality that students are only Bachelor graduates (pe=2).
Participating educators also had more concrete solutions such as collaboration in the implementation of the program (pe=3), stronger integration of content and modules across semesters (pe=3), and providing information about the content of each module across educators (pe=3).

\textit{``What you can basically say at this point is that we have enough room in the curriculum for these skills to develop, and we give people the opportunity and have many projects where they can also be honed in practice. [...] But on the other hand, it has to be said that we have taken a very ambitious approach here. Due to the interdisciplinary nature of the program, we now have various disciplines that need to be represented here. We have pretty high expectations of what people should be able to do afterward, and I think that's a good thing. The fact that we don't yet have any experience of how people will actually cope with the curriculum means that I think it's good for now and not very good.'' [PE4]}

Non-participating educators emphasized the opportunity to develop a holistic view (npe=3) and highlighted the challenge of small cohorts in the context of developing content for the specialization (npe=3 vs. pe=1).
Non-participating educators also stated multiple risks such as integration (npe=6 vs. pe=1), overload for the students (npe=3), the implementation of the study program, especially with respect to the integration of multiple modules in reality (npe=2), and missing depth in the content (npe=2). 

Overall, educators also agreed in their assessment of multiple aspects. 
An example was the assessment of interdisciplinarity, where the linking and integration of several disciplines was seen as an opportunity (npe=2 vs. np=3), employability (npe=1, pe=1).
Similarly, diversity (npe=1 vs. pe=2), and the risk of too much effort for educators was mentioned by both groups (npe=1, pe=1).
Additionally, educators from both groups emphasized communication in different professional languages (npe=1, pe=1) and interdisciplinary skills (npe=2 vs pe=3) as an expectation.

\paragraph{Differences between Industry and Educators' Perspective}
Overall, the program was perceived positively by participants from industry, indicating a good fit in terms of employability and practical aspects, as well as highlighting the relevance of the program.

\textit{``With this degree program, you really almost exactly reflect what we also expect from our employees. We would probably say that the training isn't over yet, but that someone will probably have to specialize in something [...]. But the topics that are covered are great. And when I look at things like manufacturing, production, and logistics, for example, if the people also have an idea of simulation methods, a bit of what happens in a factory or machine tools, that's really practical in any case and really an everyday toolbox for my work.'' [I2]}

To understand the differences between educators and industry, participants were asked what expectations each group had for an AI engineering graduate. Numbers are given in absolute terms, but note that the groups are unequal (14 educators, 5 industry participants).
Both groups expected professional competencies in AI concepts and technologies as well as data (educators (e)=7, industry (i)=5) and strong problem solving skills (e=8, i=4).
In addition, both groups emphasized expectations for communication across disciplines (e=2, i=1) and a holistic view of data and AI processes (e=2, i=2).

\textit{``My expectation of the AI engineer is that they are the first point of contact for potential AI projects to see whether they can be implemented and they also go through a company and see potential themselves where AI could make progress. [An AI engineer] can plan, coordinate, and support AI projects. He is also able to do a bit of prototyping. What I don't explicitly expect from an AI engineer is that they are already developing a commercial solution. [...] So for me, it's really about prototyping, dealing with methods, proof of concepts, and then actually supporting the AI project in collaboration with specialists from the respective disciplines.'' [PE5]
}

\textit{``I think due to the strong practical orientation [...] I would also expect them to have really internalized this process idea, for example, this AI engineering process from problem understanding to implementation and tracking. That's something I wouldn't expect from normal computer science students.'' [PE4]
}

Educators emphasized interdisciplinary understanding, collaboration, and acting at the intersection between domains (e=5), while industry participants expected graduates to have gained practical and project experience (i=2) and to be able to act as a mediator between domains (i=2).
For example, participants highlighted the expectation that students
\textit{``bring sufficient expertise from their application domain, understand the data that is generated, and then know how to develop customized AI solutions.'' [NPE6]}
Similarly, an expectation is that \textit{``they will ask the right, burning questions: on the basis of the data, on the basis of the information he collects, simply being able to look behind the data. I wouldn't necessarily expect them to have the full domain knowledge [...] but he can handle the data and [...] filter and derive the models that can then be used in a stringent, comprehensible, and argumentatively plausible manner.'' [I3]
}

At the same time, it was mentioned that the students were only undergraduates (n=2) and that more in-depth and advanced studies might be needed.

\textit{``So perhaps the first expectation would be that I don't expect too much. I mean, they've done a Bachelor's degree and know the vocabulary, they've done a few projects and you can definitely put them into projects with a certain amount of responsibility.'' [I1]
}

Overall, industry participants were more focused on practicality, as indicated in this interview snippet:
\textit{``[...] my experience with students is often that the tasks they are given during their studies work. The data you get there, you know that in the end, it will somehow be possible to program an AI with it. But the reality is often different. And having experienced this challenge yourself as a task - how do I get the data I need to program my AI in the end - is something that could perhaps be quite helpful.'' [I4]}

Similar to educators, industry participants also criticized the lack of content (i=6, educators=7), but the industry focus was more on practical implementations and deployments.

\textit{``[...] AI only works by putting it into practice. Purely theoretical AI is pointless. And that's why one of the points of criticism or questions for me is how much of the computer science part is also practical here. [...] how do you really build a meaningful architecture, a maintainable architecture, the compromise between perfect theory and reality, what can you really implement, where are the limits, where not.'' [I4]
}

\section{Discussion}
\label{sec:discussion}
\subsection{Main Findings}

Following the call for more curricular scholarship~\cite{krause2022vectors} and curricular frameworks for interdisciplinary AI curricula~\cite{chiu2024future}, this study contributes understanding interdisciplinary AI curriculum development with a case at the intersection of AI and engineering.
In addition to validating that the developed curriculum builds out the targeted program-level competencies, our main finding concerns the perception of educators and industry experts.

\paragraph{The developed AI and engineering curriculum is expected to be effective.}
Answering \textit{RQ1}, the curriculum mapping results indicated that the evaluated curriculum provides coverage of the targeted competencies and are built out in a coherent structure. 
In addition and following \textit{RQ2}, the focus group interviews supported these findings, with educators and industry participants generally rating the curriculum positively in terms of its fit with the competency profile.
Although educators and industry participants identified different areas for improvement, the focus group interviews supported these findings, with educators and industry participants generally rating the curriculum positively in terms of its fit with the competency profile.
In particular, participants appreciated the practical and project-oriented nature of the program, as well as its interdisciplinary nature, which is consistent with the goal of making students employable and able to work across domains.
However, the interviews also identified potential missing content (e.g., deployment, operations, pre-processing, generative AI, and foundations of computer science).
In addition, potential challenges in deepening and applying competencies such as programming skills and interdisciplinary integration were identified, indicating some potential mismatches between what is taught and how well students are able to apply these skills.

\paragraph{Participation of multiple stakeholders support the ownership of the program.}
Personal factors significantly influence curriculum choices and, thus the perceptions of quality \cite{stark_faculty_1988}. 
Thus, it could be expected that participation in curriculum development would lead to different views of outcomes. 
When comparing participating and non-participating educators (\textit{RQ3}), participation in curriculum development was found to foster a greater sense of ownership and understanding in the participating group, which positively influenced their perceptions of the program.
This implies that participatory approaches to curriculum development should be included in the design, especially for interdisciplinary programs. 
It also highlights the diversity of perspectives that underscores the value of involving diverse stakeholders in the development process.

\paragraph{Educators perceive interdisciplinarity as a strength but find it challenging to implement on a curricular level.}
Another related finding concerns interdisciplinarity. 
While there is a call for more interdisciplinary curricula and courses in engineering education~\cite{van2020interdisciplinary} and AI education~\cite{bond2024meta,chiu2024future}, the design of these is more complex than discipline-specific courses.
Overall, participants saw interdisciplinary of the program as strength, highlighting opportunities for students to gain holistic views, communicate across disciplines, and transfer skills between domains.
However, similar to previous research on integrated curriculum design~\cite{anderson2013overarching}, the integration of content across disciplines was also perceived as a challenge regarding student overload and the difficulty of effectively integrating content across disciplines and across the curriculum.
Although participants felt that the AI curriculum was well aligned with the intended interdisciplinary competency profile, they highlighted the difficulty of effectively teaching certain competencies and achieving the necessary depth given the curriculum load.
This suggests that while the curriculum is designed with the right elements, its effectiveness may be limited by time and resource constraints, especially in the context of integrating multiple domains.

This connects to previous findings of \textcite{macleod2020scaffolding}, who highlighted that scaffolding interdisciplinarity is particularly difficult in Bachelor courses, which are often taught to a broader audience of students from multiple study programs.
Thus, integration across disciplines requires more precise scaffolding for students, consistent with the learning theory of constructivism~\cite{steffe1995constructivism}, which emphasizes the importance of learners actively constructing knowledge by connecting new information to their existing understanding.
Another direction is a target use of modules or projects that integrate or communicate how competencies are combined in the curriculum~\cite{navarro2016development}.
Here, educators have a strong responsibility to choose appropriate problem designs to scaffold interdisciplinarity~\cite{macleod2020scaffolding}.
When institutions lack the resources to completely redesign and explicitly teach all modules for a program, they must find a balance between existing modules and modules that focus on transfer and integration across disciplines. 
However, more empirical work is needed to understand the conditions and interventions that support the implementation of scaffolding integration across the curriculum, in particular with a focus on AI curricula. 

\paragraph{Educators need horizontal and vertical communication across the curriculum.}
Another secondary finding of the interviews was the importance of vertical and horizontal lines of communication across the curriculum, especially for interdisciplinary programs.
This means communication between educators teaching modules in the same semester (vertical) and educators building on each other's module teaching in subsequent semesters (horizontal). 
This connects again to the aspect of 
Connecting to previous findings on integrative curriculum design~\cite{anderson2013overarching}, educators in this case study expressed that, especially in these interdisciplinary settings, it is more relevant for them to understand what has been taught and what formulations have been introduced. 
Further studies should include a better understanding of these vertical and horizontal lines of communication and how they affect the quality of student outcomes.

\paragraph{Reference curriculum for AI in engineering.} 
In addition to advancing the understanding of interdisciplinary curriculum development, the study also provides educators and curriculum developers with a practical reference curriculum for an AI in engineering program.
This can be adopted in different contexts and provides a first step towards curriculum scholarship on AI curricula in different domains~\cite{krause2022vectors,chiu2024future}.

\subsection{Limitations}
The following limitations should be considered when interpreting the results. 
First, the findings are context-specific and happen in a certain socio-cultural context that influence the overall perception of the program. 
In addition, this study focused on the conceptual development, neglecting aspects on how the curriculum of the study program is implemented and attained by students.
On the methodological side, a limitation is that the curriculum mapping was conducted based on module descriptions.
This design decision ensures reproducible results but does not take into account implicit information, validation, or context of the educators of the modules.
Another limitation is a possible selection bias in the focus group interviews conducted with a subset of educators that all gave their observations based on their experiences, which might be influenced by personal factors of educators, such as their beliefs about education or the view of their own discipline~\cite{stark_faculty_1988}.
To reduce bias in the focus groups, we focused on having a diverse group of educators and industry participants providing a diverse range of opinions. 

\subsection{Future Research Directions}
There are three main directions for future work.
First, to conduct further evaluation measures to include more student perspectives to assess the actual practicality and effectiveness of the AI curriculum.
Second, the developed curriculum should be further compared with similar developments in order to propose design principles for the development of interdisciplinary AI curricula in different contexts.
Third, the implications of the results of this study need to be further analyzed.
For example, more empirical evidence on interventions such as horizontal and vertical communication among the educators or student communication for scaffolding interdisciplinarity is needed.

\section{Conclusion}
\label{sec:conclusion}

We evaluated the development of an interdisciplinary curriculum development at the intersection of AI and engineering using a formative evaluation based on curriculum mapping and expert reviews in focus groups with educators and industry.
The study demonstrated that the developed interdisciplinary AI curriculum is expected to be effective, practical and positively validated by educators and industry.
Next to the evaluation, the study offered a practical understanding of curriculum development outcomes for AI, especially highlighting to impact of educators' participation in the development.
The curriculum profile and development approach can serve as a reference for similar initiatives at other institutions aiming at integrating AI education into disciplines.

\printbibliography

\section*{Declaration of generative AI and AI-assisted technologies in the writing process}
During the preparation of this work the authors used Grammarly and DeepL in order to improve the grammar and clarity of the text. After using this tool, the authors reviewed and edited the content as needed and take full responsibility for the content of the published article.

\section*{Ethics Statement}
All procedures were performed in compliance with relevant laws and institutional guidelines.
Informed consent was obtained from all participants involved in the study.
The participants understood their right to access and delete parts of the data, and the corresponding author anonymized the data and stored it in secured places.

\section*{CRediT Statement}
\textbf{Johannes Schleiss}: Conceptualization, Data curation, Methodology, Formal analysis, Writing – original draft, Visualization, Writing – review and editing;
\textbf{Anke Manukjan}: Data curation, Methodology, Formal analysis, Writing – review and editing;
\textbf{Michelle Ines Bieber} Data curation, Methodology, Formal analysis, Writing – review and editing;
\textbf{Sebastian Lang}: Funding acquisition, Supervision, Project administration
\textbf{Sebastian Stober}: Funding acquisition, Supervision, Project administration

\newpage
\setcounter{page}{1}
\setcounter{table}{0}
\renewcommand{\thetable}{A\arabic{table}}
\setcounter{figure}{0}
\renewcommand{\thefigure}{A\arabic{figure}}

\fancyhead[L]{\textit{Appendix: Designing an AI Curriculum for Engineering}} 
\fancyhead[C]{}                            
\fancyhead[R]{\thepage} 

\fancyfoot[L]{}                    
\fancyfoot[C]{}                            
\fancyfoot[R]{}                            
\appendix
\section{Details of the Case Study}

\subsection{Competency Profile}
\begin{table}[!htb]
	\centering
	\caption{Program level competency profile for the study program of AI Engineering}
	\small
	\begin{tabular}{p{0.15\textwidth} p{0.78\textwidth}}
		\toprule
		\textbf{Area} & \textbf{Competencies} \\
		\midrule
		Math & Understand, use, and apply mathematical methods, especially methods needed for AI systems (linear algebra, statistical and stochastic tools, analysis tools, optimization algorithms, machine learning problems) \\
		\hline
		Engineering & 
		Apply physics and engineering principles to design and develop data-driven solutions \newline
		Use measurement and sensor technology to collect and analyze engineering data \newline
		Master control engineering, signal processing techniques, and measurement and sensor technology to collect, analyze, and work with data \newline
		Design and develop engineering systems that integrate AI and data analytics \\
		\hline
		Computer \newline science & 
		Design and develop software applications using object-oriented programming \newline
		Implement algorithms and data structures to solve complex problems in AI systems\newline
		Apply software design patterns and practices to develop maintainable and scalable software \newline
		Deploy and manage cloud applications and servers \\
		\hline
		AI & 
		Collect, organize, and preprocess data to prepare it for AI model development \newline
		Develop and optimize AI models using libraries and optimization algorithms \newline 
		Visualize and evaluate AI model performance using relevant metrics and validation methods \newline
		Explain and interpret AI model results and their implications \newline
		Deploy and monitor AI systems in real-world applications \newline
		Apply agile development, continuous delivery, Machine Learning Operations (MLOps) practices to develop and deploy AI systems\\ 
		\hline
		Engineering \newline specialization & 
		Understand the fundamentals of the application area and its data-driven aspects \newline
		Apply data-driven solutions to specific engineering domains \newline
		Understand the fundamentals of the respective application area and its data-driven aspects \newline
		Map processes, data, and framework conditions in the application area to AI systems \newline
		Develop and implement AI solutions that integrate with existing engineering systems  \\
		\hline
		Process~\& system-oriented work with AI systems & 
		Develop and structure complex problems to be solved using AI systems \newline
		Plan, implement, and monitor the use of AI systems in a project \newline
		Perform value-risk assessments to ensure effective and responsible AI system deployment \newline
		Develop and implement AI systems that integrate with existing business processes and systems\\
		\hline
		Transversal skills & 
		Communicate effectively with stakeholders from diverse backgrounds and disciplines \newline
		Work collaboratively in interdisciplinary teams to develop and deploy AI systems \newline
		Plan, implement, and coordinate projects that integrate AI and data analytics \newline 
		Reflect critically on actions and skills, and adapt quickly to new topics and technologies \\
		\bottomrule
	\end{tabular}
	\label{tab:competence_profile_ai_eng}
\end{table}

\newpage
\subsection{Curricular Details of all Specializations}\label{appendix:curricular_details}
\FloatBarrier
\begin{figure}[h]
    \centering
    \includegraphics[width=0.8\linewidth]{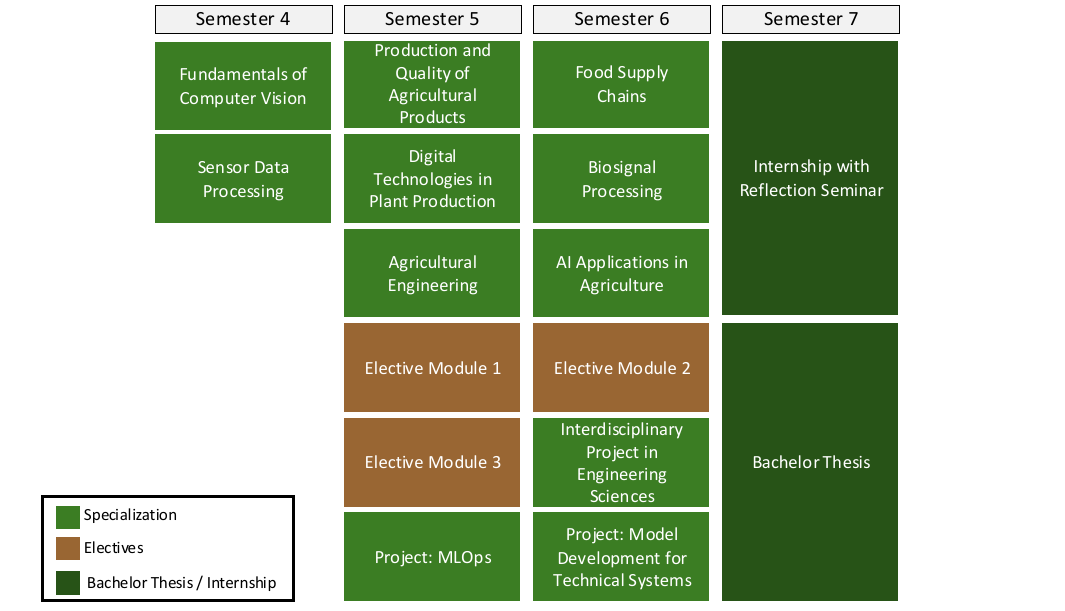}
    \caption{Curriculum of Specialization Agricultural Economy and Technology.}
    \label{fig:appendix_curriculum_specialization_1}
\end{figure}

\begin{figure}[h]
    \centering
    \includegraphics[width=0.8\linewidth]{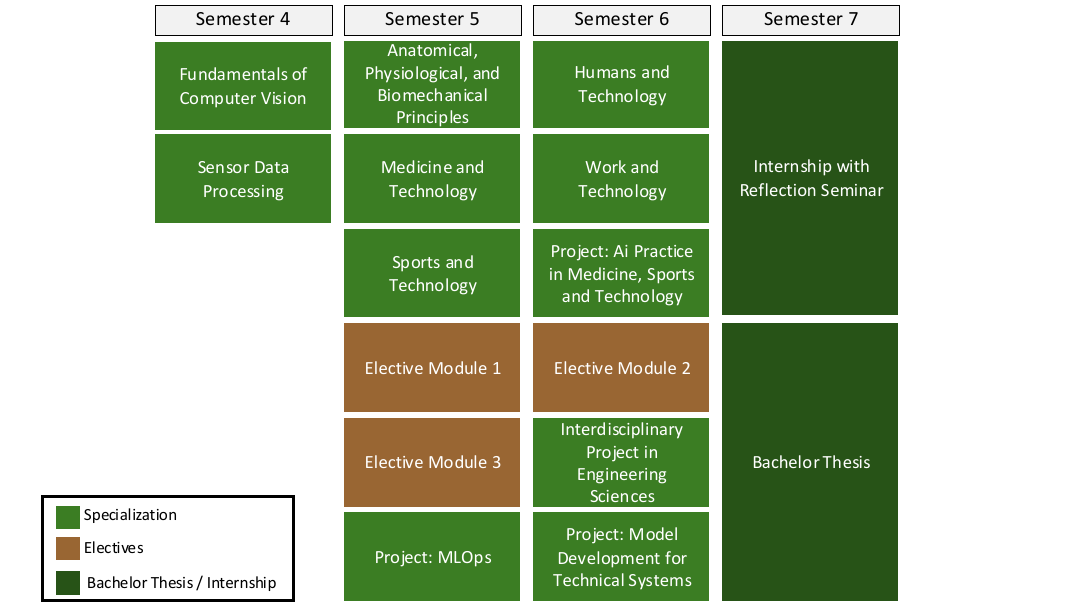}
    \caption{Curriculum of Specialization Biomechanics and Smart Health Technologies.}
    \label{fig:appendix_curriculum_specialization_2}
\end{figure}

\begin{figure}[h]
    \centering
    \includegraphics[width=0.8\linewidth]{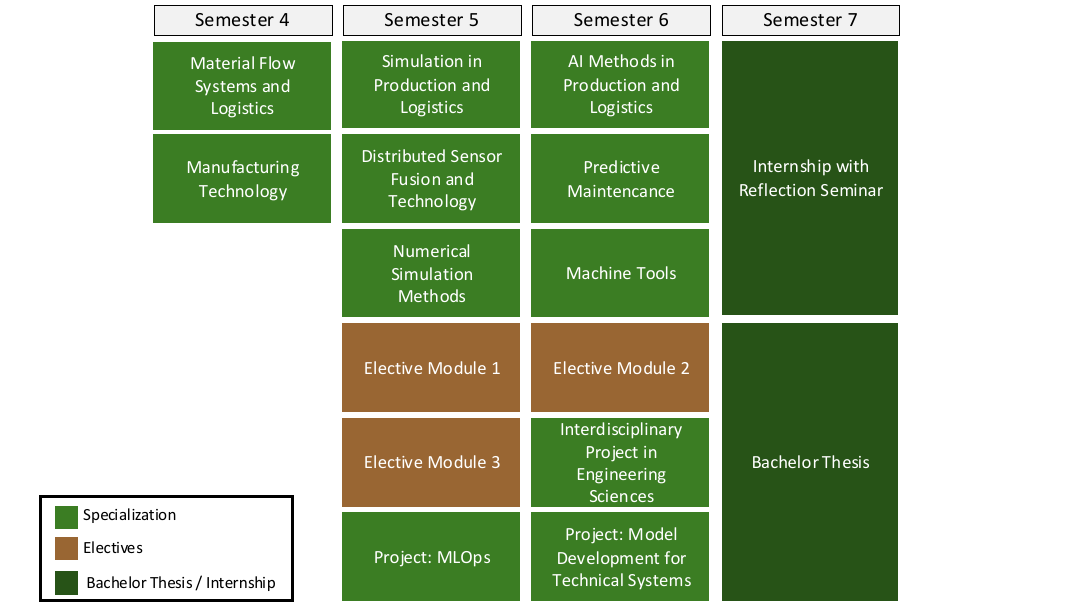}
    \caption{Curriculum of Specialization Green Engineering.}
    \label{fig:appendix_curriculum_specialization_3}
\end{figure}

\begin{figure}[h]
    \centering
    \includegraphics[width=0.8\linewidth]{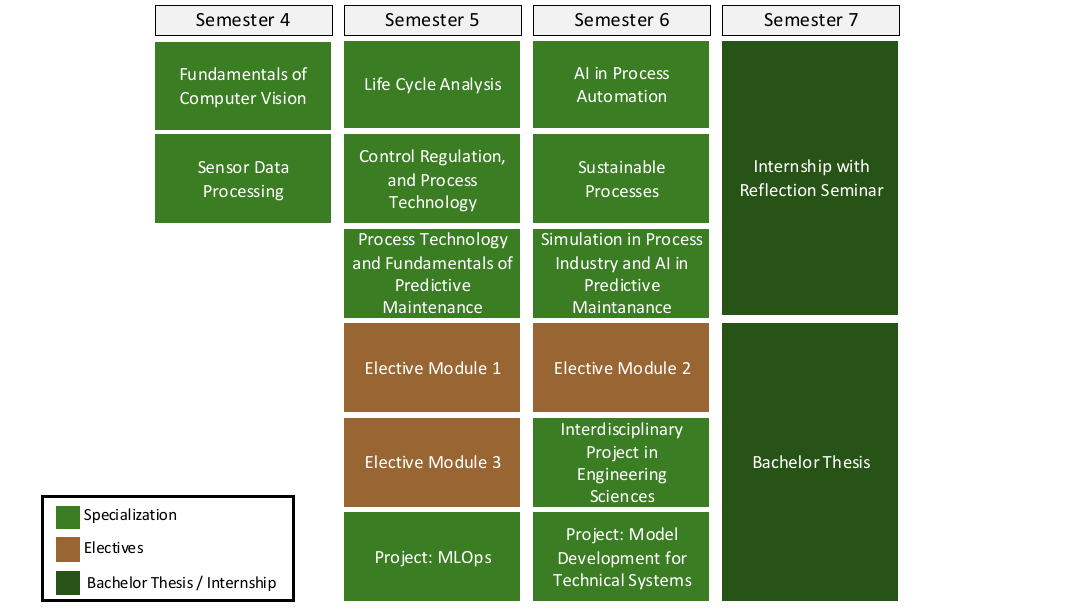}
    \caption{Curriculum of Specialization Manufacturing, Production
and Logistics.}
    \label{fig:appendix_curriculum_specialization_4}
\end{figure}
\FloatBarrier
\begin{figure}[h]
    \centering
    \includegraphics[width=0.8\linewidth]{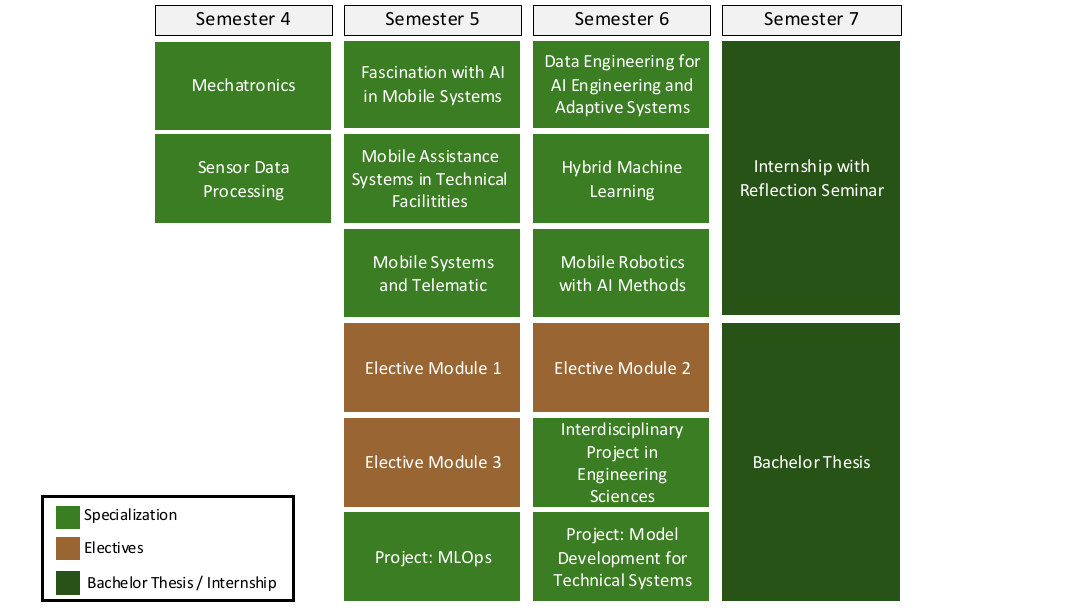}
    \caption{Curriculum of Specialization Mobile Systems and Telematics.}
    \label{fig:appendix_curriculum_specialization_5}
\end{figure}

\vspace{2cm}
\section{Curriculum Mapping Details}
\label{appendix_modulmapping}
The following \Cref{tab:curriculum_mapping} provides the complete curriculum mapping as described in \Cref{sec:methods}. In particular, it maps the competencies described in \Cref{tab:competence_profile_ai_eng} to the different modules of the curriculum indicating full \small{\fc}, partial \small{\hc}, and no coverage of subcompetencies.

\begin{landscape}
	
	\tiny
	\begin{longtable}{p{0.005\textwidth}p{0.22\textwidth} |uuuuuu | uuuuuuu|uuuu|uuuuuuu|uu|uuu|uuuuu}
		\caption[Curriculum outcome mapping.]{Curriculum outcome mapping with full, partial, and no coverage of competencies.}
		\label{tab:curriculum_mapping}\\
			
		\toprule
		ID &
		Title &
		\multicolumn{6}{l}{Math} & 
		\multicolumn{7}{l}{Engineering} &
		\multicolumn{4}{l}{CS} &
		\multicolumn{7}{l}{AI} &
		\multicolumn{2}{l}{Spec} &
		\multicolumn{3}{l}{P \& S} &
		\multicolumn{4}{l}{Transdisc} \\ \midrule
		&
		&
		\sph{Linear Algebra} &
		\sph{Statistical Tools} &
		\sph{Analysis} &
		\sph{Optimization} &
		\sph{Machine Learning} &
		\sph{Others} &
		\sph{Engineering Fundamentals} &
		\sph{Physics Fundamentals} &
		\sph{Electrical Engineering Fundamentals} &
		\sph{Measurement and Sensor Tech} &
		\sph{Applied Mechanics} &
		\sph{Control Technology} &
		\sph{Signal Processing} &
		\sph{Object-Oriented Programming} &
		\sph{Software Engineering} &
		\sph{Software Workflows} &
		\sph{Server and Cloud Handling} &
		\sph{Data Management} &
		\sph{Model Development} &
		\sph{Model Visualization} &
		\sph{Model Evaluation} &
		\sph{Ethics and Technology} &
		\sph{Agile Development} &
		\sph{AI System Monitoring} &
		\sph{Application Area Fundamentals} &
		\sph{Process and Data Analysis} &
		\sph{Problem Structuring} &
		\sph{AI Project Management} &
		\sph{Risk Assessment} &
		\sph{Cross-Disciplinary Communication} &
		\sph{Self-Organization} &
		\sph{Project Coordination} &
		\sph{Critical Reflection} \\
		\midrule
		\endfirsthead
		
		\caption[]{\textit{continued from previous page}}\\
		\toprule
		ID &
		Title &
		\multicolumn{6}{l}{Math} & 
		\multicolumn{7}{l}{Engineering} &
		\multicolumn{4}{l}{CS} &
		\multicolumn{7}{l}{AI} &
		\multicolumn{2}{l}{Spec} &
		\multicolumn{3}{l}{P \& S} &
		\multicolumn{4}{l}{Transdisc} \\ \midrule
		&
		&
		\sph{Linear Algebra} &
		\sph{Statistical Tools} &
		\sph{Analysis} &
		\sph{Optimization} &
		\sph{Machine Learning} &
		\sph{Others} &
		\sph{Engineering Fundamentals} &
		\sph{Physics Fundamentals} &
		\sph{Electrical Engineering Fundamentals} &
		\sph{Measurement and Sensor Tech} &
		\sph{Applied Mechanics} &
		\sph{Control Technology} &
		\sph{Signal Processing} &
		\sph{Object-Oriented Programming} &
		\sph{Software Engineering} &
		\sph{Software Workflows} &
		\sph{Server and Cloud Handling} &
		\sph{Data Management} &
		\sph{Model Development} &
		\sph{Model Visualization} &
		\sph{Model Evaluation} &
		\sph{Ethics and Technology} &
		\sph{Agile Development} &
		\sph{AI System Monitoring} &
		\sph{Application Area Fundamentals} &
		\sph{Process and Data Analysis} &
		\sph{Problem Structuring} &
		\sph{AI Project Management} &
		\sph{Risk Assessment} &
		\sph{Cross-Disciplinary Communication} &
		\sph{Self-Organization} &
		\sph{Project Coordination} &
		\sph{Critical Reflection} \\
		\midrule
		
		\endhead
		
		\multicolumn{36}{c}{{\textit{Continued on next page}}} \\
		\endfoot
		
		\bottomrule
		\endlastfoot
		
		1&
		Data Engineering &
		&
		&
		&
		&
		&
		&
		&
		&
		&
		&
		&
		&
		&
		&
		&
		&
		&
		\fc&
		&
		&
		&
		&
		&
		&
		&
		&
		&
		&
		&
		&
		&
		&
		\\
		2 &
		Introduction to Computer Science   for Engineers &
		&
		&
		&
		&
		&
		&
		&
		&
		&
		&
		&
		&
		&
		\fc&
		&
		&
		&
		&
		&
		&
		&
		&
		&
		&
		&
		&
		&
		&
		&
		&
		&
		&
		\\
		3 &
		Introduction to AI Engineering &
		&
		&
		&
		&
		&
		&
		&
		&
		&
		&
		&
		&
		&
		&
		&
		&
		&
		&
		&
		&
		\hc &
		\hc &
		&
		&
		&
		&
		\hc &
		\hc &
		&
		&
		&
		&
		\\
		4 &
		Electrical Engineering   Fundamentals &
		&
		&
		&
		&
		&
		&
		&
		&
		\fc&
		&
		&
		&
		\hc &
		&
		&
		&
		&
		&
		&
		&
		&
		&
		&
		&
		&
		&
		&
		&
		&
		&
		&
		&
		\\
		5 &
		Mathematics 1 for Engineers &
		\fc&
		&
		\hc &
		&
		&
		&
		&
		&
		&
		&
		&
		&
		&
		&
		&
		&
		&
		&
		&
		&
		&
		&
		&
		&
		&
		&
		&
		&
		&
		&
		&
		&
		\\
		6 &
		Engineering Design Graphics &
		&
		&
		&
		&
		&
		&
		\hc &
		&
		&
		&
		&
		&
		&
		&
		&
		&
		&
		&
		&
		&
		&
		&
		&
		&
		&
		&
		&
		&
		&
		&
		&
		&
		\\
		7 &
		Business Administration for Engineers &
		&
		&
		&
		&
		&
		&
		&
		&
		&
		&
		&
		&
		&
		&
		&
		&
		&
		&
		&
		&
		\hc &
		\hc &
		&
		&
		&
		&
		&
		&
		\hc &
		&
		&
		&
		\\
		8 &
		Fundamentals of Machine Learning &
		&
		&
		&
		&
		\hc &
		&
		&
		&
		&
		&
		&
		&
		&
		&
		&
		&
		&
		\hc &
		\hc &
		\hc &
		&
		&
		&
		&
		&
		&
		&
		&
		&
		&
		&
		&
		\\
		9 &
		Measurement technology &
		&
		&
		&
		&
		&
		&
		&
		&
		&
		\fc&
		&
		&
		&
		&
		&
		&
		&
		&
		&
		&
		&
		&
		&
		&
		&
		&
		&
		&
		&
		&
		&
		&
		\\
		10 &
		Project: Protoytping of AI-Systems &
		&
		&
		&
		&
		&
		&
		&
		&
		&
		&
		&
		&
		&
		&
		&
		&
		&
		\fc&
		\hc &
		\hc &
		\hc &
		&
		&
		\hc &
		&
		&
		\hc &
		\hc &
		&
		\hc &
		\hc &
		\hc &
		\\
		11&
		Engineering Mechanics 1&
		&
		&
		&
		&
		&
		&
		\hc &
		\hc &
		&
		&
		\hc &
		&
		&
		&
		&
		&
		&
		&
		&
		&
		&
		&
		&
		&
		&
		&
		&
		&
		&
		&
		&
		&
		\\
		12 &
		Deep Learning for Engineers &
		&
		&
		&
		&
		&
		\hc &
		&
		&
		&
		&
		&
		&
		&
		&
		&
		&
		&
		\hc &
		\fc&
		\hc &
		\fc&
		&
		&
		&
		&
		&
		&
		&
		&
		&
		&
		&
		\\
		13 &
		Industrial AI Systems &
		&
		&
		&
		&
		&
		&
		\hc &
		&
		&
		&
		&
		&
		&
		&
		&
		&
		&
		&
		&
		&
		&
		&
		&
		&
		&
		\hc &
		\fc&
		&
		&
		\hc &
		&
		&
		\\
		14 &
		Mathematics 2 for Engineers &
		&
		&
		\fc&
		&
		&
		&
		&
		&
		&
		&
		&
		&
		&
		&
		&
		&
		&
		&
		&
		&
		&
		&
		&
		&
		&
		&
		&
		&
		&
		&
		&
		&
		\\
		15 &
		Mathematics 3 for Engineers &
		&
		\fc&
		\hc &
		&
		&
		&
		&
		&
		&
		&
		&
		&
		&
		&
		&
		&
		&
		&
		&
		&
		&
		&
		&
		&
		&
		&
		&
		&
		&
		&
		&
		&
		\\
		16 &
		Project: ML Programming &
		&
		&
		&
		&
		&
		&
		&
		&
		&
		&
		&
		&
		&
		&
		&
		\hc &
		&
		\fc &
		\fc &
		\fc &
		\fc &
		&
		\hc &
		&
		&
		&
		\hc &
		\hc &
		\hc &
		\hc &
		\fc&
		\fc&
		\\
		17 &
		Software Engineering \& IT-PM &
		&
		&
		&
		&
		&
		&
		&
		&
		&
		&
		&
		&
		&
		&
		\fc&
		\hc &
		&
		&
		&
		&
		&
		&
		&
		&
		&
		&
		&
		&
		&
		&
		&
		&
		\\
		18 &
		Engineering Mechanics 2/3 &
		&
		&
		&
		&
		&
		&
		\hc &
		\hc &
		&
		&
		\fc&
		&
		&
		&
		&
		&
		&
		&
		&
		&
		&
		&
		&
		&
		&
		&
		&
		&
		&
		&
		&
		&
		\\
		19 &
		Explainable and safe AI &
		&
		&
		&
		&
		&
		&
		&
		&
		&
		&
		&
		&
		&
		&
		&
		&
		&
		&
		&
		\fc&
		\hc &
		\hc &
		&
		&
		&
		&
		&
		&
		\hc &
		&
		&
		&
		\\
		20 &
		AI based Control and Optimization   of Technical Systems and Processes &
		&
		&
		&
		\hc &
		&
		&
		\hc &
		&
		&
		&
		&
		\hc &
		&
		&
		&
		&
		&
		&
		\hc &
		&
		&
		&
		&
		&
		&
		&
		&
		&
		&
		&
		&
		&
		\\
		21&
		AI Reflection and Ethics &
		&
		&
		&
		&
		&
		&
		&
		&
		&
		&
		&
		&
		&
		&
		&
		&
		&
		&
		&
		&
		&
		\fc&
		&
		&
		&
		&
		&
		&
		\fc&
		&
		&
		&
		\hc \\
		22 &
		Signal Processing &
		&
		&
		&
		&
		&
		&
		&
		&
		&
		\hc &
		&
		&
		\fc&
		&
		&
		&
		&
		&
		&
		&
		&
		&
		&
		&
		\hc &
		&
		&
		&
		&
		&
		&
		&
		\\
		\midrule
		23 &
		Sensor Data Processing &
		&
		&
		&
		&
		&
		&
		&
		&
		&
		&
		&
		\fc&
		&
		&
		&
		&
		&
		&
		&
		&
		&
		&
		&
		&
		\hc &
		&
		&
		&
		&
		&
		&
		&
		\\
		24 &
		Introduction to Computer Vision &
		&
		&
		&
		&
		&
		&
		&
		&
		&
		&
		&
		&
		&
		&
		&
		&
		&
		&
		&
		&
		&
		&
		&
		&
		\hc &
		&
		&
		&
		&
		\hc &
		\hc &
		&
		\\
		25 &
		Digital technologies in   agriculture production &
		&
		&
		&
		&
		&
		&
		&
		&
		&
		&
		&
		&
		&
		&
		&
		&
		&
		&
		&
		&
		&
		&
		&
		&
		\hc &
		\hc &
		\hc &
		&
		&
		&
		&
		&
		\\
		26 &
		Production and Quality of   Agricultural Products &
		&
		&
		&
		&
		&
		&
		&
		&
		&
		&
		&
		&
		&
		&
		&
		&
		&
		&
		&
		&
		&
		&
		&
		&
		\hc &
		&
		&
		&
		&
		&
		&
		&
		\\
		27 &
		Agricultural Technology &
		&
		&
		&
		&
		&
		&
		&
		&
		&
		&
		&
		&
		&
		&
		&
		&
		&
		&
		&
		&
		&
		&
		&
		&
		\hc &
		\hc &
		&
		&
		&
		&
		&
		&
		\\
		28 &
		Biosignal Processing &
		&
		&
		&
		&
		&
		&
		&
		&
		&
		&
		&
		&
		&
		&
		&
		&
		&
		\hc &
		&
		&
		&
		&
		&
		&
		\hc &
		\hc &
		&
		&
		&
		&
		&
		&
		\\
		29 &
		AI Applications in Agriculture &
		&
		&
		&
		&
		&
		&
		&
		&
		&
		&
		&
		&
		&
		&
		&
		&
		&
		&
		&
		&
		&
		&
		&
		&
		\hc &
		\hc &
		&
		&
		&
		&
		&
		&
		\\
		30 &
		Food Supply Chains &
		&
		&
		&
		&
		&
		&
		&
		&
		&
		&
		&
		&
		&
		&
		&
		&
		&
		&
		&
		&
		&
		&
		&
		&
		\hc &
		&
		&
		&
		&
		&
		&
		&
		\\
		31&
		Project MLOps &
		&
		&
		&
		&
		&
		&
		&
		&
		&
		&
		&
		&
		&
		&
		&
		&
		\fc&
		&
		&
		&
		&
		&
		\fc&
		\fc&
		&
		&
		&
		\hc &
		&
		\fc&
		\fc&
		\fc&
		\\
		32 &
		Project Model development for   technical systems &
		&
		&
		&
		&
		&
		&
		&
		&
		&
		&
		&
		&
		&
		&
		&
		&
		&
		\fc&
		\fc&
		\fc&
		\fc&
		&
		&
		&
		&
		&
		\fc&
		\fc&
		&
		\fc&
		\hc &
		&
		\hc \\
		33 &
		Interdisciplinary Project &
		&
		&
		&
		&
		&
		&
		&
		&
		&
		&
		&
		&
		&
		&
		&
		&
		&
		\hc &
		\hc &
		\hc &
		\hc &
		&
		\hc &
		&
		\hc &
		\hc &
		\hc &
		\hc &
		\hc &
		\fc&
		\hc &
		&
		\hc \\
		\midrule
		23 &
		Sensor Data Processing &
		&
		&
		&
		&
		&
		&
		&
		&
		&
		&
		&
		\fc&
		&
		&
		&
		&
		&
		&
		&
		&
		&
		&
		&
		&
		\hc &
		&
		&
		&
		&
		&
		&
		&
		\\
		24 &
		Introduction to Computer Vision &
		&
		&
		&
		&
		&
		&
		&
		&
		&
		&
		&
		&
		&
		&
		&
		&
		&
		&
		&
		&
		&
		&
		&
		&
		\hc &
		&
		&
		&
		&
		\hc &
		\hc &
		&
		\\
		34 &
		Introduction to medical   engineering &
		&
		&
		&
		&
		&
		&
		&
		&
		&
		&
		&
		&
		&
		&
		&
		&
		&
		&
		&
		&
		&
		&
		&
		&
		&
		&
		&
		&
		&
		&
		&
		&
		\\
		35 &
		Introduction to sports science and sports engineering &
		&
		&
		&
		&
		&
		&
		&
		&
		&
		&
		&
		&
		&
		&
		&
		&
		&
		&
		\hc &
		&
		&
		&
		&
		&
		\hc &
		\hc &
		\hc &
		&
		&
		&
		&
		&
		\\
		36 &
		Work and technology - AI &
		&
		&
		&
		&
		&
		&
		&
		&
		&
		&
		&
		&
		&
		&
		&
		&
		&
		&
		&
		&
		&
		\hc &
		&
		&
		\hc &
		\hc &
		&
		&
		&
		&
		&
		&
		\\
		37 &
		Introduction to human anatomy, physiology and biomechanics &
		&
		&
		&
		&
		&
		&
		&
		&
		&
		&
		&
		&
		&
		&
		&
		&
		&
		&
		&
		&
		&
		&
		&
		&
		\hc &
		&
		&
		&
		&
		&
		&
		&
		\\
		38 &
		Humans and technology &
		&
		&
		&
		&
		&
		&
		&
		&
		&
		&
		&
		&
		&
		&
		&
		&
		&
		&
		&
		&
		&
		\hc &
		&
		&
		&
		&
		&
		&
		&
		&
		&
		&
		\hc \\
		39 &
		Project: The practical application of AI in medicine, sports and technology &
		&
		&
		&
		&
		&
		&
		&
		&
		&
		&
		&
		&
		&
		&
		&
		&
		&
		\hc &
		\hc &
		&
		&
		&
		&
		&
		\hc &
		\hc &
		\hc &
		\hc &
		&
		&
		&
		&
		\\
		31&
		Project MLOps &
		&
		&
		&
		&
		&
		&
		&
		&
		&
		&
		&
		&
		&
		&
		&
		&
		\fc&
		&
		&
		&
		&
		&
		\fc&
		\fc&
		&
		&
		&
		\hc &
		&
		\fc&
		\fc&
		\fc&
		\\
		32 &
		Project Model development for technical systems &
		&
		&
		&
		&
		&
		&
		&
		&
		&
		&
		&
		&
		&
		&
		&
		&
		&
		\fc&
		\fc&
		\fc&
		\fc&
		&
		&
		&
		&
		&
		\fc&
		\fc&
		&
		\fc&
		\hc &
		&
		\hc \\
		33 &
		Interdisciplinary Project &
		&
		&
		&
		&
		&
		&
		&
		&
		&
		&
		&
		&
		&
		&
		&
		&
		&
		\hc &
		\hc &
		\hc &
		\hc &
		&
		\hc &
		&
		\hc &
		\hc &
		\hc &
		\hc &
		\hc &
		\fc&
		\hc &
		&
		\hc \\
		\midrule
		40 &
		Manufacturing Processes 1&
		&
		&
		&
		&
		&
		&
		&
		&
		&
		&
		\fc&
		&
		&
		&
		&
		&
		&
		&
		&
		&
		&
		&
		&
		&
		\hc &
		&
		&
		&
		&
		&
		&
		&
		\\
		41&
		Material Handling Systems and Logistics &
		&
		&
		&
		&
		&
		&
		\hc &
		&
		&
		&
		&
		&
		&
		&
		&
		&
		&
		&
		&
		&
		&
		&
		&
		&
		\hc &
		&
		&
		&
		&
		&
		&
		&
		\\
		42 &
		Numerical methods for simulation &
		&
		&
		\hc &
		\hc &
		&
		&
		&
		&
		&
		&
		&
		&
		&
		&
		&
		&
		&
		&
		&
		&
		&
		&
		&
		&
		\hc &
		&
		&
		&
		&
		&
		&
		&
		\\
		43 &
		Simulation in Production and Logistik &
		&
		&
		&
		&
		&
		&
		&
		&
		&
		&
		&
		&
		&
		&
		&
		&
		&
		&
		&
		&
		&
		&
		&
		&
		\hc &
		&
		&
		&
		&
		&
		&
		&
		\\
		44 &
		Introduction to Distributed Sensor   Data Fusion &
		&
		\hc &
		&
		&
		&
		&
		&
		&
		&
		\fc&
		&
		&
		\hc &
		&
		&
		&
		&
		&
		&
		&
		&
		&
		&
		&
		\hc &
		&
		&
		&
		&
		&
		&
		&
		\\
		45 &
		AI methods in production planning and control &
		&
		&
		&
		&
		&
		&
		\hc &
		&
		&
		&
		&
		&
		&
		&
		&
		&
		&
		&
		\hc &
		\hc &
		\hc &
		&
		&
		&
		\hc &
		\hc &
		\hc &
		&
		&
		&
		&
		&
		\\
		46 &
		Predictive Maintenance &
		&
		&
		&
		&
		&
		&
		\hc &
		&
		&
		&
		&
		&
		&
		&
		&
		&
		&
		&
		\hc &
		\hc &
		&
		&
		&
		&
		\hc &
		\hc &
		\hc &
		&
		&
		&
		&
		&
		\\
		47 &
		Machine Tools &
		&
		&
		&
		&
		&
		&
		\hc &
		&
		&
		&
		\hc &
		&
		&
		&
		&
		&
		&
		&
		&
		&
		&
		&
		&
		&
		\hc &
		&
		&
		&
		&
		&
		&
		&
		\\
		31&
		Project MLOps &
		&
		&
		&
		&
		&
		&
		&
		&
		&
		&
		&
		&
		&
		&
		&
		&
		\fc&
		&
		&
		&
		&
		&
		\fc&
		\fc&
		&
		&
		&
		\hc &
		&
		\fc&
		\fc&
		\fc&
		\\
		32 &
		Project Model development for   technical systems &
		&
		&
		&
		&
		&
		&
		&
		&
		&
		&
		&
		&
		&
		&
		&
		&
		&
		\fc&
		\fc&
		\fc&
		\fc&
		&
		&
		&
		&
		&
		\fc&
		\fc&
		&
		\fc&
		\hc &
		&
		\hc \\
		33 &
		Interdisciplinary Project &
		&
		&
		&
		&
		&
		&
		&
		&
		&
		&
		&
		&
		&
		&
		&
		&
		&
		\hc &
		\hc &
		\hc &
		\hc &
		&
		\hc &
		&
		\hc &
		\hc &
		\hc &
		\hc &
		\hc &
		\fc&
		\hc &
		&
		\hc \\
		\midrule
		23 &
		Sensor Data Processing &
		&
		&
		&
		&
		&
		&
		&
		&
		&
		&
		&
		\fc&
		&
		&
		&
		&
		&
		&
		&
		&
		&
		&
		&
		&
		\hc &
		&
		&
		&
		&
		&
		&
		&
		\\
		24 &
		Introduction to Computer Vision &
		&
		&
		&
		&
		&
		&
		&
		&
		&
		&
		&
		&
		&
		&
		&
		&
		&
		&
		&
		&
		&
		&
		&
		&
		\hc &
		&
		&
		&
		&
		\hc &
		\hc &
		&
		\\
		48 &
		Life cycle analysis &
		&
		&
		&
		&
		&
		&
		&
		&
		&
		&
		&
		&
		&
		&
		&
		&
		&
		&
		&
		&
		&
		&
		&
		&
		\hc &
		&
		&
		&
		&
		\hc &
		&
		&
		\\
		49 &
		Process technology and   fundamentals of predictive maintenance &
		&
		&
		&
		&
		&
		&
		&
		&
		&
		&
		&
		&
		&
		&
		&
		&
		&
		&
		&
		&
		&
		&
		&
		&
		\hc &
		\hc &
		&
		&
		&
		&
		&
		&
		\\
		50 &
		Control-, regulation- and process   control engineering &
		&
		&
		&
		&
		&
		&
		&
		&
		&
		&
		&
		&
		&
		&
		&
		&
		&
		&
		&
		&
		&
		&
		&
		&
		&
		&
		&
		&
		&
		&
		&
		&
		\\
		51&
		AI in Process Automation &
		&
		&
		&
		&
		&
		&
		&
		&
		&
		&
		&
		&
		&
		&
		&
		&
		&
		&
		\hc &
		\hc &
		&
		&
		&
		&
		\hc &
		\hc &
		&
		&
		&
		&
		&
		&
		\\
		52 &
		Sustainable processes &
		&
		&
		&
		&
		&
		&
		&
		&
		&
		&
		&
		&
		&
		&
		&
		&
		&
		&
		&
		&
		&
		&
		&
		&
		\hc &
		&
		&
		&
		&
		&
		&
		&
		\\
		53 &
		Simulation in the Process Industry   and AI in Predictive Maintenance &
		&
		&
		&
		\hc &
		&
		&
		&
		&
		&
		&
		&
		&
		&
		&
		&
		&
		&
		&
		&
		&
		&
		&
		&
		&
		\hc &
		\hc &
		&
		&
		&
		&
		&
		&
		\\
		31&
		Project MLOps &
		&
		&
		&
		&
		&
		&
		&
		&
		&
		&
		&
		&
		&
		&
		&
		&
		\fc&
		&
		&
		&
		&
		&
		\fc&
		\fc&
		&
		&
		&
		\hc &
		&
		\fc&
		\fc&
		\fc&
		\\
		32 &
		Project Model development for   technical systems &
		&
		&
		&
		&
		&
		&
		&
		&
		&
		&
		&
		&
		&
		&
		&
		&
		&
		\fc&
		\fc&
		\fc&
		\fc&
		&
		&
		&
		&
		&
		\fc&
		\fc&
		&
		\fc&
		\hc &
		&
		\hc \\
		33 &
		Interdisciplinary Project &
		&
		&
		&
		&
		&
		&
		&
		&
		&
		&
		&
		&
		&
		&
		&
		&
		&
		\hc &
		\hc &
		\hc &
		\hc &
		&
		\hc &
		&
		\hc &
		\hc &
		\hc &
		\hc &
		\hc &
		\fc&
		\hc &
		&
		\hc \\
		\midrule
		54 &
		Mechatronics 1&
		&
		&
		&
		&
		&
		&
		&
		&
		&
		&
		&
		\hc &
		&
		&
		&
		&
		&
		&
		&
		&
		&
		&
		&
		&
		\hc &
		&
		&
		&
		&
		&
		&
		&
		\\
		55 &
		Sensor Data Processing &
		&
		&
		&
		&
		&
		&
		&
		&
		&
		&
		&
		\fc&
		&
		&
		&
		&
		&
		&
		&
		&
		&
		&
		&
		&
		\hc &
		&
		&
		&
		&
		&
		&
		&
		\\
		56 &
		Fascination AI in mobile systems &
		&
		&
		&
		&
		&
		&
		&
		&
		&
		&
		&
		&
		&
		&
		&
		&
		&
		&
		&
		&
		&
		&
		&
		&
		\hc &
		\hc &
		&
		&
		&
		&
		&
		&
		\\
		57 & Mobile assistance systems in the operation of technical systems
		&
		&
		&
		&
		&
		&
		&
		&
		&
		&
		&
		&
		&
		&
		&
		&
		&
		&
		\hc &
		&
		&
		&
		&
		&
		&
		\hc &
		\hc &
		\hc &
		\hc &
		&
		&
		&
		&
		\\
		58 &
		Mobile systems \& telematics &
		&
		&
		&
		&
		&
		&
		&
		&
		&
		&
		&
		&
		&
		&
		&
		&
		&
		\hc &
		&
		&
		&
		&
		&
		&
		\hc &
		\hc &
		&
		&
		&
		&
		&
		&
		\\
		59 &
		Data Engineering and Adaptive   Intelligent Systems &
		&
		&
		&
		&
		&
		&
		&
		&
		&
		&
		&
		&
		&
		&
		&
		&
		&
		\hc &
		&
		&
		&
		&
		&
		&
		\hc &
		&
		&
		&
		&
		&
		&
		&
		\\
		60 &
		Hybrid machine learning -   knowledge and data-based models of technical systems &
		&
		&
		&
		&
		&
		&
		&
		&
		&
		&
		&
		&
		&
		&
		&
		&
		&
		&
		&
		&
		&
		&
		&
		&
		\hc &
		\hc &
		&
		&
		&
		&
		&
		&
		\\
		61&
		Mobile robotics with AI methods &
		&
		&
		&
		&
		&
		&
		&
		&
		&
		&
		&
		&
		&
		&
		&
		&
		&
		&
		&
		&
		&
		&
		&
		&
		\hc &
		\hc &
		&
		&
		&
		&
		&
		&
		\\
		31&
		Project MLOps &
		&
		&
		&
		&
		&
		&
		&
		&
		&
		&
		&
		&
		&
		&
		&
		&
		\fc&
		&
		&
		&
		&
		&
		\fc&
		\fc&
		&
		&
		&
		\hc &
		&
		\fc&
		\fc&
		\fc&
		\\
		32 &
		Project Model development for technical systems &
		&
		&
		&
		&
		&
		&
		&
		&
		&
		&
		&
		&
		&
		&
		&
		&
		&
		\fc&
		\fc&
		\fc&
		\fc&
		&
		&
		&
		&
		&
		\fc&
		\fc&
		&
		\fc&
		\hc &
		&
		\hc \\
		33 &
		Interdisciplinary Project &
		&
		&
		&
		&
		&
		&
		&
		&
		&
		&
		&
		&
		&
		&
		&
		&
		&
		\hc &
		\hc &
		\hc &
		\hc &
		&
		\hc &
		&
		\hc &
		\hc &
		\hc &
		\hc &
		\hc &
		\fc&
		\hc &
		&
		\hc
	\end{longtable}
\end{landscape}

\newpage
\section{Questionnaire for Focus Group Interview}
\label{appendix:interview_guide}
\normalsize

\subsubsection*{Introduction}
The introduction followed a standard procedure. 
First, the moderators introduced themselves and presented the agenda and objectives with a emphasis on the evaluation of the design of the program not the implementation.
Next, focus groups as a method was introduced and participants introduced themselves with name and professional background. 
Last, the tools were introduced and final consent for recording obtained. 

\subsubsection*{Participant Backgrounds via Online Tool}
\textbf{1.1 Teaching experience}: How long have you been teaching?

\textbf{1.2 Curriculum development experience}: Have you ever contributed to developing a curriculum?

\textbf{1.3 Attitude towards interdisciplinary programs}: What is your attitude towards interdisciplinary programs in general?

\subsubsection*{Topic: Representation of the Competency Profile through the Curriculum}
We would like to give you five minutes to review the competency profile (which we have already sent to you in advance). Afterwards, we will present the module matrix and would like to answer the following question (see 2.1).

\textbf{2.1 Is the competency profile achieved with this curricular arrangement of modules?} (Number and justification). (Scale: \textit{1 - very good} to \textit{5 - not good at all})

\textbf{Discussion}: Justification of the evaluation, inquiries, and deepening in the insights and reasoning

\subsubsection*{Topic: Content, Expectations, and Acceptance}
From a content and disciplinary perspective:

\textbf{3.1 How well does the curriculum meet the professional requirements of your discipline / your area of expertise (e.g., computer science, mechanical engineering, etc.)?} (Scale 1 Very good to 5 not good at all)

Discussion: Justification of the evaluation, inquiries, and deepening in the insights and reasoning

\textbf{3.2 Strengths and weaknesses}: Where do you see strengths and weaknesses in the curriculum?

\textbf{3.3 Structure}:  How do you assess the structure of the program in the proposed course sequence? Do the subjects build meaningfully on each other?

\textbf{3.4 Expectations}: What expectations do you have of an AI Engineer?

\subsubsection*{Topic: Interdisciplinarity}
\textbf{4.1 Interdisciplinarity}: In relation to the program: Where do you see opportunities and risks in interdisciplinarity?

\subsubsection*{Optional Additional Questions}
\textbf{O.1 Studyability}: How appropriate do you find the content and workload for the target audience?

\textbf{O.2 Changes}: What changes would you recommend for the curriculum to better prepare students for the challenges and opportunities that may arise in the future?

\textbf{O.3 Balance}: How do you assess the balance between foundational knowledge and specialization?

\subsubsection*{Conclusion and Thanks}
Are there any further comments from your side?  
Thank you and next steps: Brief questionnaire – also for additional ideas that may arise

\section{Coding Scheme}
\label{appendix_coding_scheme}
\vspace{-0.5cm}
\begin{table}[h]
\centering
\caption{Coding scheme per interview part and coding themes and subthemes with codes occurring more than one time.}
\label{tab:coding_scheme}
\footnotesize
\begin{tabular}{p{2.7cm} p{2.5cm} p{7.7cm}}
\toprule
\textbf{Interview Part}    & \textbf{Theme} &\textbf{Subtheme} \\
\midrule
\multirow{16}{2.5cm}{General evaluation of the study program} & \multirow{5}{2.5cm}{Strengths} & interdisciplinary orientation \\
 &  & practical/project-orientation \\
 &  & diversity of content \\
 &  & curriculum \\
 &  & employability \\ \cline{2-3} 
 & \multirow{2}{2.5cm}{Weakness} & more innovation possible \\
 &  & employability \\ \cline{2-3} 
 & \multirow{5}{2.5cm}{Challenges} & lack of fundamentals/content \\
 &  & small cohorts \\
 &  & provide orientation \\
 &  & realization \\
 &  & practical/project-orientation \\ \cline{2-3} 
 & \multirow{3}{2.5cm}{Possible solutions}& Cooperation in implementation of study program \\
 &  & information about the content of the modules \\
 &  & increased interlinking \\ \cline{2-3} 
 & Improvements & identification of core practices in the field of AI engineering \\ 
\midrule
\multirow{16}{2.5cm}{Interdisciplinarity} & \multirow{8}{2.5cm}{Opportunities} & attractiveness \\
 &  & interlinking \\
 &  & practical/project-orientation \\
 &  & versatility \\
 &  & change of perspective \\
 &  & profiling \\
 &  & holistic view \\
 &  & communication/ technical language \\ \cline{2-3} 
 & \multirow{5}{2.5cm}{Risks} & overload \\
 &  & interlinking \\
 &  & implementation \\
 &  & effort \\
 &  & lack of depth \\ \cline{2-3} 
 & \multirow{3}{2.5cm}{Challenges} & cooperation during implementation \\
 &  & communication \\
 &  & interlinking \\
\midrule
\multirow{7}{2.5cm}{Expectations\\ of an AI Engineer} & Expectations & interdisciplinary working and understanding \\
 &  & professional competencies \\
 &  & communication in different technical languages \\
 &  & practical/project experience \\
 &  & understanding of the overall process \\
 &  & problem-solving strategies \\
 &  & mediating role \\
\midrule
\multirow{10}{2.5cm}{Curriculum and competence profile fit} & Remarks & confirmation \\ \cline{2-3}  & \multirow{6}{2.5cm}{Challenges} & provide orientation \\
 &  & representing transversal skills \\
 &  & implementation \\
 &  & deepening and application of competences \\
 &  & interlinking \\
 &  & cooperation during implementation \\ \cline{2-3} 
 & Criticism & missing fundamentals/content \\ \cline{2-3} 
 & Positive & practical/project-orientation \\ \cline{2-3} 
 & Risks & overload\\

 \bottomrule
\end{tabular}%
\end{table}
\FloatBarrier
\end{document}